\documentclass[%
 reprint,
 amsmath,amssymb,
 aps,
 floatfix,
]{revtex4-2}

\usepackage[utf8]{inputenc}
\usepackage[T1]{fontenc}
\usepackage{microtype}
\usepackage{graphicx}
\usepackage{bm}
\usepackage{textcomp}
\usepackage{xcolor}
\usepackage{xurl}
\usepackage{booktabs}
\usepackage{multirow}
\usepackage{subcaption}
\usepackage{hyperref}
\hypersetup{hidelinks}
\usepackage[capitalise,nameinlink]{cleveref}

\begin{document}

\preprint{SQUIRO}

\title{SQUIRO: A Framework for Security-Aware Quantum-Classical Scheduling on Kubernetes}

\author{Ignazio Pedone}
\email{ignazio@helix-ft.com}
\affiliation{Helix 42, Rome, Italy}

\author{Edoardo Giusto}
\email{egiusto@ieee.org}
\affiliation{Helix 42, Rome, Italy}
\affiliation{University of Naples Federico II, Naples, Italy}

\date{\today}

\begin{abstract}
Distributed infrastructure schedulers traditionally optimise capacity, locality, and cost, but provide limited support for security posture and emerging quantum-classical workloads.
As hybrid quantum-classical computing becomes increasingly practical and post-quantum security requirements begin to affect infrastructure deployment, schedulers must jointly reason about heterogeneous compute resources, security constraints, and quantum backend characteristics.
We present SQUIRO, a framework for security-aware quantum-classical scheduling based on a platform-independent Unified Scheduling Model (USM) and a six-step Scheduler Design Methodology (SDM) that together enable  the derivation of concrete schedulers for Kubernetes, high-performance computing (HPC), and federated environments.
The framework combines multidimensional security posture enforcement through hard feasibility constraints with residual-risk optimisation, and introduces a circuit-aware quantum backend selector that accounts for coherence margin, calibration freshness, queue pressure, and hardware capabilities through a forward-compatible colocation hierarchy.
Evaluation on synthetic Kubernetes clusters shows that the security model enforces complete compliance for regulated workloads by construction, while global optimisation reduces infrastructure cost by up to 51\% and energy consumption by up to 63\% compared with greedy placement in underloaded scenarios, without compromising admission priorities.
Additional experiments characterise the solve-time growth of the current CP-SAT formulation and show that circuit-aware backend selection systematically diverges from naive error-rate ranking under coherence- and queue-limited conditions.
\end{abstract}


\maketitle

\section{Introduction}
\label{sec:introduction}
Schedulers for distributed infrastructures fundamentally rely on factors such as capacity and locality, while operating with limited visibility into workload and infrastructure models.
This restriction is primarily driven by efficiency: richer models induce hard optimisation problems whose solution times often exceed the operational timescales of the managed system.
In large-scale clusters, however, such simplifications can lead to inefficient resource utilisation, suboptimal exploitation of heterogeneous hardware, and avoidable security risks.
Security posture is typically absent from allocation decisions despite becoming an increasingly critical concern.
AI-driven attacks and the quantum threat, together with the ongoing post-quantum cryptography (PQC) migration~\cite{nistPqcMigration}, exemplify practical pressures arising from the rapid expansion of discovered vulnerabilities, attack surfaces, and threat classes that current systems are not yet prepared to address~\cite{fang2024llmagents,
anthropic2025gtg1002,giusto2025typology}.
This is also a scheduling-time concern: NIST-standardised post-quantum key encapsulation and signature algorithms~\cite{nistPqcStandards} are already being deployed, compliance mandates are taking effect~\cite{omb2026m2615}, and nodes that cannot satisfy post-quantum requirements should be excluded from regulated workload placement before any objective function is evaluated.

Distributed infrastructures are also becoming increasingly heterogeneous, with quantum computing providing a representative example.
Quantum systems are now accessible as cloud services, while the algorithms that use them are inherently hybrid.
A variational eigensolver, a quantum-accelerated optimisation routine, or a quantum machine learning pipeline follows the same basic pattern: a classical program prepares the input, submits a gate-level circuit to a Quantum Processing Unit (QPU) for execution, and uses the returned measurement results to determine the next classical step.
This loop may be repeated thousands of times until a convergence criterion is met.

\begin{table*}[htbp]
\centering
\caption{Comparison of SQUIRO vs other scheduling systems grouped by
topic. \checkmark\,=\,integrated; $\circ$\,=\,partial or
limited; ---\,=\,absent.}
\label{tab:comparison}
\small
\setlength{\tabcolsep}{3.5pt}
\renewcommand{\arraystretch}{1.05}
\begin{tabular}{@{}llccccc@{}}
\toprule
\textbf{Topic} & \textbf{System} & \textbf{Kubernetes} &
\textbf{QC scheduling} & \textbf{Security risk} &
\textbf{Cost/energy} & \textbf{Formal model} \\
\midrule

\multirow{4}{*}{Infrastructure}
& Kubernetes native \cite{kubernetesSchedulingFramework} & \checkmark &
--- & --- & --- & --- \\
& Gavel \cite{narayanan2020heterogeneity} & --- & --- & --- &
\checkmark & $\circ$ \\
& Quincy \cite{isard2009quincy} & --- & --- & --- & --- & \checkmark \\
& Firmament \cite{gog2016firmament} & --- & --- & --- & --- &
\checkmark \\
\midrule

\multirow{3}{*}{QC schedulers}
& Qurator \cite{pehlivanoglu2026qurator} & --- & \checkmark & ---
& --- & --- \\
& SCIM~MILQ \cite{seitz2024scimmilq} & --- & \checkmark & ---
& --- & $\circ$ \\
& Qonductor \cite{giortamis2025qonductor} & \checkmark & \checkmark
& --- & --- & --- \\
\midrule

\multirow{2}{*}{Security-aware}
& Al-Haj et al.~\cite{alhaj2013security} & --- & --- &
$\circ$ & --- & \checkmark \\
& Felemban et al.~\cite{felemban2025riskaware} & --- & --- & $\circ$
& --- & $\circ$ \\
\midrule

Formal basis
& Classical scheduling \cite{pinedo2016scheduling} & --- & --- & ---
& --- & \checkmark \\

\midrule
& SQUIRO (this work) & \checkmark & \checkmark & \checkmark
& \checkmark & \checkmark \\
\bottomrule
\end{tabular}
\end{table*}


Hybrid quantum-classical execution therefore requires two coupled scheduling choices.
First, the scheduler must determine where the classical stages execute, selecting a node that provides sufficient capacity while satisfying the workload security requirements.
Second, it must choose an appropriate quantum backend whose coherence time, connectivity, and queueing delay are compatible with the quantum workload. 
Classical node placement determines the data-locality relationship with the selected backend.
A mismatch, such as placing a data-intensive preprocessing stage on a node that is geographically, administratively, or network-wise distant from the backend, can itself violate the loop timing constraints.
Conventional infrastructure schedulers do not address such decisions jointly.

Standard cloud orchestration platforms schedule workloads through local filter-and-score heuristics, typically considering one workload at a time. They provide no explicit representation of quantum resources, no security-risk model, and no mechanism for globally minimising cost or energy across the pending workload set. Quantum middleware systems address workflow- and backend-level scheduling, but often require explicit backend selection and are commonly tied to specific software stacks, implementations, or runtimes. Security-aware placement research has embedded compliance constraints into classical allocation models; however, these approaches do not address PQC readiness, trusted-execution evidence, or the threat surface specific to multi-tenant quantum execution, where co-scheduled workloads may infer circuit structure or execution timing from shared hardware state~\cite{erata2024quantum}.

Quantum-classical scheduling can also be viewed as a \textit{three-level scheduling problem} that must ultimately be co-designed, but can currently only be loosely coordinated. Circuit compilers and transpilers~\cite{tejedor2025qdislib} operate 
at the top
layer:
they decompose quantum algorithms into hardware-native gate sequences, and their output determines which backends are eligible and at what expected fidelity.
Quantum abstraction layers at the operating system (OS) level~\cite{ramsauer2025towards} and Kubernetes device plugins~\cite{kubernetesDevicePlugins} operate 
at the bottom
layer: they mediate direct QPU access and expose calibration state, thereby bounding the execution guarantees that any infrastructure-level placement decision can actually deliver. The infrastructure layer sits between the previous two and is responsible for node placement, backend selection, and distributed resource allocation. This paper focuses on this layer.

To address security posture and quantum-computing awareness at scheduling time, this paper presents SQUIRO, a \textbf{S}ecurity-aware \textbf{Qu}antum-classical \textbf{I}nfrastructure for \textbf{R}esource-scheduling and \textbf{O}rchestration.
SQUIRO is built around two contributions.
First, it introduces the platform-independent Unified Scheduling Model (USM) and the accompanying six-step Scheduler Design Methodology (SDM), which together provide a domain-independent formalisation of scheduling problems and a systematic derivation process that maps this abstraction to concrete schedulers by defining the deployment scope, control architecture, scheduling algorithm, and implementation data model.
Because the approach is platform independent, the same derivation applies to cloud clusters, HPC batch facilities, and federated hybrid infrastructures.
Second, it designs and implements SQUIRO as a Kubernetes scheduler, combining a multi-objective optimisation model that jointly accounts for monetary cost, energy consumption, and security posture with quantum-aware on-demand backend selection.

Experiments on synthetic clusters show an initial advantage over a Kubernetes-like greedy scheduler in terms of cost and energy optimisation, as well as security posture handling. On the quantum side, circuit-aware backend selection further demonstrates that ranking backends by raw error rate alone is insufficient: circuit depth and execution constraints can shift the optimal choice in ways that a naive selector cannot capture. 


The rest of the paper is organised as follows.
Section~\ref{sec:background} reviews related work. Sections~\ref{sec:model}--\ref{sec:squiro_framework} develop the model, methodology, and framework. Section~\ref{sec:application} details the Kubernetes instantiation. Sections~\ref{sec:evaluation}--\ref{sec:conclusion} present the preliminary evaluation and conclusions.

\section{Background and Related Work}
\label{sec:background}

The following subsections will detail how SQUIRO compares against other scheduling systems present in the literature across five dimensions, summarised in Table~\ref{tab:comparison}.
The comparison is at the
scheduler-capability level: each entry reflects whether the system
integrates the capability into a unified scheduling decision rather than
treating it as a separate concern.

\subsection{Classical Scheduling Theory}
\label{subsec:classical_theory}

Pinedo's
framework \cite{pinedo2016scheduling} is
the standard formalism for machine scheduling:
it classifies the machine environment, the job characteristics and constraints, and the objective.


Gavel \cite{narayanan2020heterogeneity} shows that
accelerator-aware placement requires explicit performance modelling per
device type; Quincy \cite{isard2009quincy} and Firmament
\cite{gog2016firmament} show that global min-cost flow over a resource
graph improves placement quality.  
These systems and models do not
extend to typed quantum resources, execution-mode-dependent backend
constraints, or security posture as a first-class scheduling dimension.
SQUIRO generalises Pinedo's framework by adopting USM, detailed in \Cref{sec:model}.
The resource graph \(G_R\) replaces
the flat machine environment with typed edges encoding trust boundaries and quantum access paths; the workload descriptor adds capability requirements and execution semantics. Moreover, the policy layer $\mathcal{P}$, which governs constraints and objectives, has no analogue in Pinedo's formalism.

\subsection{Quantum-HPC and Cloud Integration}
\label{subsec:quantum_hpc}

Early quantum-HPC infrastructure work established the need for middleware
above raw device access. XACC \cite{mccaskey2020xacc} introduced
heterogeneous quantum-classical execution with service-oriented interfaces;
subsequent work identified the gap between HPC workload semantics and
quantum execution requirements
\cite{shehata2026bridging,beck2024integrating,saurabh2023conceptual}. Pilot-Quantum
\cite{mantha2025pilot} and follow-on adaptive middleware
\cite{mantha2026adaptive} apply the pilot-job abstraction to hybrid
workloads. Bacher et al.~\cite{bacher2025quantum} introduce plugins for
Slurm (the de facto standard workload manager for HPC batch facilities) that expose on-premises and cloud QPUs as schedulable resources within
the familiar job-submission model. The Quantum Device Management Interface (QDMI)
\cite{burgholzer2026qdmi,wille2024qdmi} standardises the device-management
boundary between HPC software and quantum backends. The openQSE survey
\cite{shehata2026openqse} identifies recurring gaps in runtime abstraction,
resource management, and orchestration.

At the cloud-native level, Qubernetes \cite{stirbu2024qubernetes}
demonstrates quantum jobs within the Kubernetes ecosystem. Qonductor
\cite{giortamis2025qonductor} extends this with a hardware-agnostic API
for hybrid workload scheduling. Tejedor et al.
\cite{tejedor2026kubernetes} coordinate CPUs, GPUs, and QPUs through
Kubernetes, Argo Workflows, and Kueue. Qunicorn \cite{weder2024qunicorn}
addresses provider heterogeneity and API fragmentation.
Pedone and Lioy \cite{pedone2022qkdkubernetes} demonstrate quantum key distribution (QKD) integration
within Kubernetes clusters, establishing a software stack for
quantum-secure key exchange in containerised environments; earlier work
\cite{pedone2021qkdcloud} discusses the same topic within the broader scope of cloud orchestration.
Among the systems reviewed here, we found no system that combines infrastructure placement with systematic security-risk assessment and joint optimisation of cost, energy, and risk.

\subsection{Quantum-Centric Architectures and Scheduling}
\label{subsec:quantum_centric}

Several quantum-centric supercomputing perspectives \cite{seelam2026qcentric,alexeev2024quantum}
describe an evolution from isolated QPU offload
to co-designed heterogeneous quantum-HPC systems coordinated across
application, middleware, and infrastructure layers.
Ramsauer and Mauerer
\cite{ramsauer2025towards} propose a Quantum Abstraction Layer at the
OS level to support low-latency interaction and generic
resource management for quantum accelerators.
These
developments indicate that QPUs may eventually appear as schedulable
node-level capabilities rather than remote provider endpoints, a transition
the SQUIRO colocation class hierarchy accommodates without
changes to the abstract model.

At the scheduling layer, variability-aware policies showed that hardware
heterogeneity should influence mapping and execution decisions
\cite{tannu2018case}. SCIM~MILQ \cite{seitz2024scimmilq} formulates HPC
quantum scheduling combining circuit cutting and classical scheduling
theory. Qurator \cite{pehlivanoglu2026qurator} generalises hybrid
quantum-cloud scheduling as dynamic DAG optimisation across providers.
Cipollini et al.~\cite{cipollini2026threeways} compare multiplexing
strategies for scarce QPU resources. Distributed circuit cutting
\cite{tejedor2025qdislib} and resource-aware circuit scheduling
\cite{choudhury2025quartet} operate at the circuit and compilation level.
Quantum data-centre scheduling \cite{yang2025qdc} confirms that QPU
clusters require explicit resource management. Dependability studies
\cite{giusto2025dependable} and quantum trusted execution environments
\cite{trochatos2025qtee} identify security and provenance as core design
concerns for hybrid quantum-HPC systems.  Across this body of
work, QPU scheduling is addressed at the circuit or device layer;
infrastructure-level allocation that jointly decides placement, backend
selection, and security posture remains an open problem.

\subsection{Security- and Risk-Aware Allocation}
\label{subsec:security_allocation}

Security-aware resource allocation has been studied in classical cloud
settings. Al-Haj et al.~\cite{alhaj2013security} formulate VM allocation
with reachability and access-control requirements as a constraint problem.
Felemban et al.~\cite{felemban2025riskaware} model multi-tenant
data-leakage risk as a hard assignment problem. Quantum clouds introduce
additional exposure channels through crosstalk and shared-device effects
\cite{arellano2025qubitvise}.  These works motivate embedding
security inside the scheduling model.
SQUIRO extends this foundation to the
quantum-classical setting: it adds PQC posture,
trusted computing evidence, distributed quantum control-plane security, and
AI model risk as first-class scheduling dimensions, and introduces a strict
separation between non-negotiable requirements and optimisable residual
risk.






\section{Unified Scheduling Model}
\label{sec:model}

\begin{figure}
\centering
\includegraphics[width=\columnwidth]{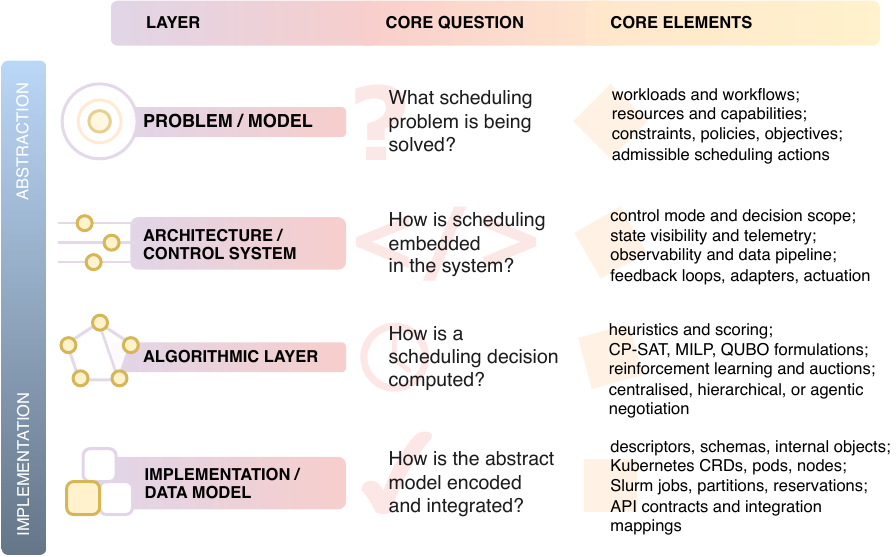}
\caption{Four-layer decomposition of the scheduling system.
The 
USM
addresses the problem/model layer, while the
SDM
traces the path
through the remaining layers to a concrete platform instantiation.}
\label{fig:layer_decomposition}
\end{figure}

This section presents the Unified Scheduling Model: a
platform-independent formal description of a scheduling system,
applicable to SQUIRO as well as to cloud, HPC, or hybrid computing
systems in general. The USM can be adapted and instantiated in different architectural contexts such as a centralised scheduler optimiser, a decentralised agentic system, or a federated approach.
The model is also flexible with respect to how the scheduling problem is solved: the designer may use a particular formulation such as mixed-integer linear programming (MILP) or quadratic unconstrained binary optimisation (QUBO) or specific solvers based on simulated annealing, quantum annealing, or custom classical heuristics.
The USM definition is the first step of a \textbf{four-layer decomposition} to characterise a general scheduling system for computing, as in Figure \ref{fig:layer_decomposition}.
The \emph{problem/model} layer  formalises what resources
and workloads exist and how feasible plans for allocation are defined.
The \emph{architecture/control} layer embeds the scheduler in an operational context,
determining what state is observable and what actions are authorised as well as
the type of architecture (e.g., centralised, decentralised).
Together, these layers determine the problem available to the algorithm.
The \emph{algorithmic} layer 
governs how scheduling decisions are computed from that observable state.
The final layer is \emph{implementation and data model}, which translates abstract
decisions into platform-specific descriptors and enforcement mechanisms.

\subsection{USM Instance Definition}
\label{subsec:instance_definition}

A scheduling instance at time \(t\) is defined as
\begin{equation}
\mathcal{I}_t =
\left(
G_W,\;
G_R,\;
S_t,\;
\mathcal{C},\;
\mathcal{P},\;
\mathcal{O},\;
\mathcal{A}
\right)
\label{eq:general_instance}
\end{equation}
where
\begin{itemize}
\item \(G_W = (W, E_W)\) is the \emph{workload graph}, i.e. a \textit{workflow}. Vertices are
executable workloads or stages; edges represent dependencies, data-flow,
synchronisation, or control-flow relations.

\item \(G_R = (R, E_R)\) is the \emph{resource graph}. Vertices are
schedulable resources or resource domains; edges represent network
connectivity, data-transfer cost, administrative boundaries, trust
boundaries, or provider access paths.

\item \(S_t\) is the \emph{scheduler-visible state} at time \(t\). It
encompasses current allocations, available capacity, queue depth, and
active reservations; cost and power signals; quantum backend calibration
snapshots; and security evidence such as attestation records and
compliance metadata. The control architecture determines how \(S_t\) is
obtained; the model assumes a state estimate is available.
\(S\) denotes the space of all such states, with \(S_t \in S\) its value at time \(t\).

\item \(\mathcal{C}\) is the set of \emph{hard constraints}. A schedule
violating any element of \(\mathcal{C}\) is infeasible regardless of its
objective value.

\item \(\mathcal{P}\) is the \emph{policy layer}. Policies express
governance rules as well as operator priorities, resource quotas, fairness
criteria, compliance mandates, and risk thresholds. They also govern how
constraints and objectives are compiled and enforced. Policies may add elements to
\(\mathcal{C}\) or modify the objective ranking.

\item \(\mathcal{O}\) is the \emph{objective model}, defining how feasible
schedules are compared. Relevant criteria include completion time,
resource cost, energy consumption, security risk exposure, and quantum
execution quality. The policy layer determines which apply and how they
combine.

\item \(\mathcal{A}\) is the \emph{admissible decision space}: the set of
actions the scheduler may output. Depending on the deployment scope and
control boundary, this encompasses placement, admission or deferral
decisions. It also includes resource reservation and backend assignment.
\end{itemize}

The scheduler is a decision process defined as
\begin{equation}
\pi_t : \mathcal{I}_t \longrightarrow \mathcal{A}
\label{eq:scheduler_mapping}
\end{equation}
that maps the current instance to an admissible action. The algorithm
implementing \(\pi_t\) is not part of the model, and, as 
discussed,
it can use different approaches.

\subsection{Workloads}
\label{subsec:workloads_workflows}

Each workload is described by a minimal descriptor
\begin{equation}
w_i = (d_i,\; K_i,\; p_i,\; \eta_i),
\label{eq:workload_descriptor}
\end{equation}
where $d_i = (d_{i1}, \ldots, d_{i|D|})$ is the resource-demand vector over a shared dimension set $D$ (e.g. compute, memory, storage, network, accelerator capacity), with $d_{ik}$ the demand of workload $i$ in dimension $k \in D$, $K_i$ the set of required capabilities, \(p_i\) a
function \(p_i : 2^R \times S \to \mathbb{R}_{\geq 0}\) giving the
processing time of workload \(i\) under a potential  allocation and state,
and \(\eta_i\) the execution semantics (e.g., batch, preemptible, elastic,
gang-scheduled).
In a cloud setting, \(w_i\) may represent a
pod, job, or service component; in an HPC setting a batch or MPI job; in a
quantum-classical setting a circuit-generation stage, quantum execution
stage, or error-mitigation step.

\subsection{Resources and Capabilities}
\label{subsec:resources_capabilities}

The resource graph is
\begin{equation}
G_R = (R, E_R), \quad R = \{r_1, \ldots, r_m\}.
\label{eq:resource_graph}
\end{equation}
Each resource \(r_j\) is described by
\begin{equation}
r_j = \bigl(\mathrm{cap}_j(t),\; K_j(t)\bigr),
\label{eq:resource_descriptor}
\end{equation}
where \(\mathrm{cap}_j(t)\) is the available capacity vector over the same
dimension set \(D\) used by \(d_i\), with \(\mathrm{cap}_{jk}(t)\) its
component in dimension \(k \in D\), and \(K_j(t)\) is the current
capability set. Capabilities absorb heterogeneity without requiring a
separate model for every technology: a capability may describe CPU
architecture, GPU type, trusted execution environment (TEE) support,
quantum-backend properties, or security posture attributes.

Edges \(E_R\) encode typed relations between resources. An
edge may represent a network link (with latency and bandwidth attributes),
a data-transfer cost, an administrative or jurisdictional boundary, a
trust boundary between security domains, or an inter-facility transfer
path between cloud and HPC resources.

\subsection{Feasibility and Objectives}
\label{subsec:feasibility_objectives}

A time-aware allocation plan is
\begin{equation}
x = (z,\; \mu,\; s,\; \rho),
\label{eq:schedule}
\end{equation}
where \(z_i \in \{0,1\}\) is the admission variable, \(\mu(w_i)
\subseteq R\) the resource allocation, \(s_i \in \mathbb{R}_{\geq 0}\)
the start time, \(\rho(e)\) an optional mapping of workflow edges to
paths in \(G_R\), and \(c_i = s_i + p_i(\mu(w_i), S_t)\) the completion
time. \(\mathcal{X}\) denotes the space of all plan tuples
$(z, \mu, s, \rho)$. The feasible plan set is
\begin{equation}
\mathcal{X}_{\mathrm{feas}}(t)
= \bigl\{x \in \mathcal{X} \mid g(x, S_t) = 1,\;
\forall g \in \mathcal{C}\bigr\},
\label{eq:feasible_set}
\end{equation}
where each \(g : \mathcal{X} \times S \to \{0,1\}\) is a constraint
predicate. The core capacity, precedence, and capability constraints are,
respectively:
\begin{align}
\sum_{\substack{i:\, z_i=1,\, r_j \in \mu(w_i) \\
s_i \leq \tau < c_i}} d_{ik} &\leq \mathrm{cap}_{jk}(\tau), \quad \forall k \in D,
\label{eq:capacity_constraint} \\
z_i = z_k = 1 \Rightarrow {s_k} &\geq c_i + {\Delta_{ik}}, \quad {(w_i, w_k)} \in E_W,
\label{eq:precedence_constraint} \\
z_i = 1 \Rightarrow K_i &\subseteq \bigcup_{r_j \in \mu(w_i)} K_j(t),
\label{eq:capability_constraint}
\end{align}
where {\(\Delta_{ik} \geq 0\)} is the data-transfer or communication delay
on edge {\((w_i, w_k)\)}. The precedence and capability constraints are
conditioned on admission because \(\mu(w_i)\) and \(c_i\) are only
meaningful once workload \(i\) is actually assigned resources: for
\(z_i=0\), no assignment exists to check, and rejection must not be
blocked by a constraint over an allocation that was never made.
Rejecting a predecessor does not exclude its successors: for \(z_i=0\), Eq.~\eqref{eq:precedence_constraint} leaves \(z_k\) unconstrained for any \((w_i,w_k) \in E_W\), so partial workflow admission is permitted by construction.
{Additional feasibility
constraints, such as node affinity, workload isolation requirements, and
attestation bounds, are added as predicates in \(\mathcal{C}\) without altering
the base model structure.} The objective model assigns a cost to each feasible plan:
\begin{equation}
x^\star \in \arg\min_{x \in \mathcal{X}_{\mathrm{feas}}(t)} F(x, S_t).
\label{eq:scalar_objective}
\end{equation}
For multi-objective settings, \(F\) is replaced by a vector
\(\mathbf{F}(x, S_t) = (F_1, \ldots, F_k)\), ranked through a
policy-governed selection rule:
\begin{equation}
x^\star \in \operatorname{Select}_{\mathcal{P}, \mathcal{O}}
\bigl(\mathcal{X}_{\mathrm{feas}}(t),\; \mathbf{F}\bigr).
\label{eq:selection_rule}
\end{equation}

\section{Scheduler Design Methodology}
\label{sec:methodology}

The proposed methodology maps requirements $\mathcal{R}$ through a preliminary model instance $\mathcal{I}_t$, an architecture choice $\mathsf{Arch}$, an induced problem class $\mathcal{Q}$, an algorithmic strategy $\mathsf{Alg}$, and an implementation environment $\mathcal{E}$:
\[
\mathcal{R} \;\to\; \mathcal{I}_t \;\to\; \mathsf{Arch} \;\to\; \mathcal{Q}
\;\to\; \mathsf{Alg} \;\to\; \mathcal{E}.
\]
This progression yields the six steps used to design a scheduling platform with the USM, making it derivable and extensible; the ordering matters, since each step constrains the valid choices in the next. The methodology proceeds as follows:

\begin{enumerate}
\item \textbf{Analyse scope, context, and objectives.}
This step fixes, for the target domain: the workload and resource scope (e.g., elastic cloud pods vs. fixed-capacity HPC batch jobs); the observable state and its acquisition path, which constrains the control architecture selected in step~3; trust assumptions across resource and administrative boundaries; the environmental baseline constraints; and the sources of inefficiency, threat, and rigidity that become the optimisation objectives in step~4.

\item \textbf{Instantiate the USM.}
The analytical output of step~1 is mapped onto the model components,
constructing $G_W$, $G_R$, $\mathcal{C}$, $\mathcal{P}$, $\mathcal{O}$, and a preliminary $S_t$ and $\mathcal{A}$ for the target scope, both finalised in step~3 once the control architecture fixes their observation and control boundary.
Domain-specific requirements, including security
constraints, backend eligibility rules, and compliance thresholds, must
enter as model components at this step: a constraint added as a
post-processing filter cannot participate in feasibility or influence the
objective.

\item \textbf{Select the control architecture.}
The choice of control architecture, whether centralised, hierarchical, federated, or decentralised, is determined by the state visibility and
authority established in step~1. This decision fixes the control boundary
and determines how state is observed and how resource management decisions propagate through the system, finalising the preliminary $S_t$ and $\mathcal{A}$ from step~2. The architecture introduces structural
trade-offs in solution quality and decision
latency that persist regardless of the algorithm selected in step~4: a
centralised design observes the full system state and enables tighter
optimisation, while decentralised or federated designs may reduce decision latency and visibility, with a possible loss of global solution quality.
Hybrid architectures
are feasible when different resource domains require different control
granularities.

\item \textbf{Derive the mathematical problem and select an algorithm.}
Based on the outputs of steps 1 through 3, the induced scheduling problem
is classified and one or more algorithmic approaches are identified. The
problem class may correspond to general allocation, multidimensional
bin-packing, job-shop scheduling, or minimum-cost flow, depending on the
structural properties of the workload and resource graphs. The choice of
algorithmic approach is not necessarily tied to the nature of the underlying
computing substrate: both classical methods, such as MILP, constraint
programming, and filter-and-score heuristics, and QUBO formulations, including those considered for quantum annealing
\cite{chiavassa2022vnfqa}, are valid candidates.
This step yields an
inventory of options with explicit trade-offs among mathematical complexity,
time to solution, and solution quality.

\item \textbf{Define the implementation and data model.}
The abstract model components are translated into concrete resource and
workload descriptors, schemas, and platform APIs, mapping abstract requirements unambiguously onto the computing resources and domain constraints of the target environment. Extensions such as finer-grained configuration, compliance metadata, or attestation evidence must not degrade scheduling performance or compromise the enforceability of constraints in \(\mathcal{C}\).

\item \textbf{Validate, iterate, and evolve.}
Validation pursues two distinct aims. The first is to surface cross-layer
interactions that are not apparent when layers are treated as independent
components. The second is to resolve
conflicts between the abstract model and practical implementation
constraints: platform or domain limitations discovered during
implementation may reduce the feasible solution space and must be handled
as special cases, either by tightening the model or by demoting
unenforceable constraints from \(\mathcal{C}\) to preferences in
\(\mathcal{O}\). A constraint without an enforcement mechanism must not be treated as a hard guarantee: the solver may produce plans the system cannot execute.
\end{enumerate}


\begin{figure*}
\centering
\includegraphics[width=\textwidth]{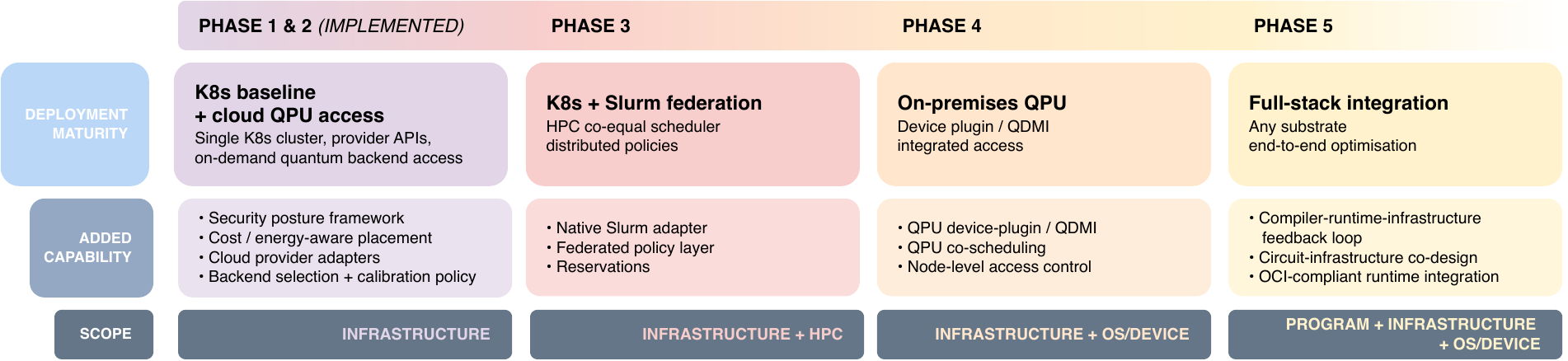}
\caption{Roadmap of the staged evolution of SQUIRO, from the already implemented Kubernetes (K8s) and cloud-QPU foundation toward full end-to-end quantum-classical stack optimisation.
}
\label{fig:roadmap}
\end{figure*}

\section{The SQUIRO Framework}
\label{sec:squiro_framework}
SQUIRO is a scheduling and orchestration framework for quantum-classical workloads that jointly optimises security alongside cost and energy.
It is designed toward an end-to-end quantum-classical scheduling pipeline in which the three scheduling layers (circuit, infrastructure, and OS/device) are coupled; full end-to-end optimisation across all layers is the long-term SQUIRO roadmap target (Phase 5), as detailed in Figure \ref{fig:roadmap}.
Through the use of USM and SDM, the framework targets infrastructure-level scheduling with a centralised control architecture as its starting point, and is designed to extend incrementally across a wider execution substrate and a deeper optimisation scope.
The current implementation
covers \textit{Phases~1--2} of the roadmap:
a centralised scheduler operating over a Kubernetes cluster with access to on-demand cloud quantum backends.
Two external modules augment the core optimiser.
The \emph{security posture framework} computes a per-node security score, derives hard security constraints, and quantifies residual risk; all three outputs feed directly into the multi-objective optimisation alongside cost and energy.
The \emph{quantum backend selector} scores available remote backends against workload compatibility requirements and provides the interface for submitting quantum workloads to cloud providers.
Together they provide inputs for the joint \textit{cost--energy--security--quality} optimisation; in the current prototype, residual risk and backend selection are not yet fully integrated into the core solver, and are instead enforced as a hard constraint and a separate callable service respectively.
\textit{Phase~3} extends the framework's deployment to HPC
environments, in particular Slurm, enabling cloud and HPC platforms to
be used jointly under the same abstract model. It additionally
introduces a decentralised variant of the architecture for
distributed scheduling.
\textit{Phase~4} adds on-premises QPU management through device
plugins or QDMI, shifting QPU resources from opaque remote services to
locally observable entities whose queue dynamics and calibration state
become visible to the scheduler. At this stage, the multi-objective
optimisation is extended to directly incorporate optimal quantum
backend selection through a dedicated term in the policy-governed
objective (Eq.~\ref{eq:squiro_scalar_objective}, \Cref{subsec:squiro_instantiation}).
\textit{Phase~5} represents the long-term end-to-end target: joint optimisation across all three scheduling layers from compile-time and transpilation decisions through runtime placement and OCI-compliant container scheduling.

The rationale for this phased progression is the inherent coupling between scheduling layers in quantum-classical systems. Circuit-level choices constrain which backends are eligible; infrastructure placement determines data locality and communication overhead; OS-level access modes bound the execution guarantees that can be offered. Optimising any single layer independently leaves inter-layer efficiency unexploited and can produce placements that are locally optimal but globally suboptimal.
SDM is the mechanism that makes this extension tractable: each new phase introduces new adapters and widens the scope definition, but the abstract model and the security and quantum frameworks remain unchanged.

\section{SQUIRO Kubernetes Integration}
\label{sec:application}
\subsection{Kubernetes Scheduling}
\label{subsec:squiro_kubernetes_context}

The default Kubernetes scheduler observes unscheduled pods, filters
infeasible nodes, scores feasible nodes, and binds a pod to the selected
node~\cite{kubernetesSchedulingFramework,kubernetesSchedulerConfig}.
While effective for throughput and simplicity, this filter-and-score loop
has two fundamental limitations in a quantum-classical context.

The first limitation is the absence of global scheduling visibility.
Deciding one pod at a time with local heuristics provides no mechanism
for cluster-wide cost minimisation, energy consolidation, or
security-aware placement.
A cost-minimising scheduler must pack demand onto the smallest active node set, a goal not achievable by construction without a global view over all pending workloads simultaneously. A security-aware scheduler must enforce placement constraints as hard requirements that cannot be traded against cost; per-pod local filtering can already enforce this, though reaching the smallest compliant node set still requires the same global view.

\begin{table*}[tbp]
\centering
\caption{SQUIRO USM-to-Kubernetes mapping.}
\label{tab:squiro_model_mapping}
\renewcommand{\arraystretch}{1.12}
\setlength{\tabcolsep}{3pt}
\begin{tabular}{@{}p{0.12\linewidth}p{0.22\linewidth}p{0.58\linewidth}@{}}
\toprule
\textbf{Model term} &
\textbf{Kubernetes interpretation} &
\textbf{Design component} \\
\midrule

\(G_W\) &
Workload graph &
Pod and Job descriptors, grouped workloads, quantum execution requests,
and optional workflow or gang relationships. \\

\(G_R\) &
Resource graph &
Nodes, zones, storage and network domains, trusted domains,
simulators, and on-demand quantum backends. \\

\(S_t\) &
Scheduler-visible state &
Current placements, allocatable capacity, labels, taints, metrics,
cost and power estimates, security evidence, and backend telemetry. \\

\(\mathcal{C}\) &
Hard feasibility constraints &
Capacity, affinity, toleration, GPU, OS, security,
locality, and backend-eligibility constraints. \\

\(\mathcal{P}\) &
Policies and preferences &
Priority, budgets, security thresholds, backend-selection preferences,
risk acceptance, and permitted actuation modes. \\

\(\mathcal{O}\) &
Objective model &
Admission reward, cost, energy, fragmentation, locality, disruption,
residual risk, queue delay, and quantum execution quality. \\

\(\mathcal{A}\) &
Admissible decisions &
Placement, admission, deferral, rejection, validation,
recommendation, backend assignment, and replanning. \\

\bottomrule
\end{tabular}
\end{table*}

The second limitation is the absence of a quantum-aware control path.
To our knowledge, no native or officially supported Kubernetes mechanism currently provides
the scheduling layer with quantum backend metadata, such as calibration state,
queue depth, coherence times, or circuit compatibility, at the point of
placement decisions. Quantum-classical workflows therefore cannot be
co-scheduled with their classical counterparts using standard Kubernetes
primitives.

SQUIRO addresses both limitations through two complementary integration
modes. In \emph{co-pilot mode}, SQUIRO reads workload and cluster state
from the live cluster, solves the global optimisation problem, and
returns placement or backend recommendations that an operator can enforce
manually or automatically according to a configurable policy, leaving
lifecycle management and reconciliation to Kubernetes. In
\emph{orchestrator mode}, workloads and resources are specified directly
via SQUIRO descriptors, giving the scheduler full control over placement
in deployments where tighter integration is preferred.
Both modes
support global batch scheduling over all pending workloads as well as
incremental scheduling over the next ready workloads, enabling
dynamic allocation without accumulating a full batch.

\subsection{Model Instantiation and Scheduler Design}
\label{subsec:squiro_instantiation}

The SDM of Section~\ref{sec:methodology} underpins this
instantiation; for compactness and to build mathematical vocabulary
progressively, the results of the derivation chain are presented in a
different order from the derivation sequence. The model mapping and
descriptors are introduced first to establish a shared vocabulary for the
problem formulation and architecture choice that follow; both are motivated
where they appear. Two extensions, the quantum backend selector and the
security posture framework, are described in the following subsections.

\subsubsection*{Model Mapping and Descriptors}

Table~\ref{tab:squiro_model_mapping} translates each model component to
its Kubernetes interpretation, followed by the descriptor definitions
that the problem formulation and architecture workflow below reference directly.
Six descriptor types normalise raw Kubernetes objects and provider
telemetry into the uniform representation required by \(\mathcal{I}_t\):

\begin{itemize}
\item \textit{Workload.} Maps to vertex \(w_i \in W\), capturing identity, namespace, grouping, and policy scope.

\item \textit{Demand.} Forms the resource request vector \(d_i\) across all scheduling dimensions (CPU, memory, storage, network, GPU), used directly in the capacity constraints.

\item \textit{Compatibility.} Encodes Kubernetes-specific feasibility requirements, such as node affinity, anti-affinity, tolerations, required GPU model, and OS, into the binary mask \(a^k_{ij}\).

\item \textit{Scheduling semantics.} Carries priority, deadline, gang identifier, and replica count, feeding the admission reward, grouped placement constraints, and the admissible-decision set \(\mathcal{A}\).

\item \textit{Node/resource.} Maps to resource vertex $r_j \in R$ with capacity vector $\mathrm{cap}_j$ and capability set $K_j(t)$, derived from CPU, memory, GPU inventory, storage, and network bandwidth.

\item \textit{Objective signals.} Supplies cost rate, power estimate, and utilisation summaries, entering the cost \(C(y)\) and energy \(E(y)\) terms and the consolidation and fragmentation penalties in \(\mathcal{O}\).
\end{itemize}

\subsubsection*{Allocation Problem}
\label{subsec:squiro_placement}

The Kubernetes instantiation operates on a scheduler input snapshot:
start times and edge realisation are delegated to Kubernetes, and the
optimiser decides admission and node placement for all candidate
workloads simultaneously. The Kubernetes-restricted plan is
\begin{equation}
x_k = (z,\; \mu),
\label{eq:k8s_plan}
\end{equation}
where \(\mu(w_i)\) is a singleton Kubernetes node or the empty set.

Let \(W\) be the candidate workloads, \(R\) the candidate nodes, and
\(D = \{\mathrm{cpu}, \mathrm{mem}, \mathrm{storage}, \mathrm{net},
\mathrm{gpu}\}\) the resource dimensions. Let \(d_{ik}\) be the demand
of workload \(i\) in dimension \(k\) and \(\mathrm{cap}_{jk}\) the available
capacity of node \(j\). Binary variables \(x_{ij}\) (workload \(i\)
placed on node \(j\)), \(y_j\) (node \(j\) active), and \(z_i\)
(workload \(i\) admitted) satisfy:
\begin{align}
\sum_{j \in R} x_{ij} &= z_i, && \forall i \in W,
\label{eq:squiro_assignment} \\
\sum_{i \in W} d_{ik} x_{ij} &\leq \mathrm{cap}_{jk}\, y_j,
&& \forall j \in R,\; \forall k \in D,
\label{eq:squiro_capacity} \\
x_{ij} &\leq y_j, && \forall i \in W,\; \forall j \in R.
\label{eq:squiro_activation}
\end{align}

Kubernetes-specific feasibility, e.g., node affinity, labels, OS, and GPU model requirements, is encoded by a
binary compatibility mask \(a^k_{ij} \in \{0,1\}\):
\begin{equation}
x_{ij} \leq a^k_{ij}, \qquad \forall i \in W,\; \forall j \in R.
\label{eq:squiro_k8s_compatibility}
\end{equation}
Gang groups \(G \subseteq W\) enforce all-or-nothing admission: any
two workloads \(i, i' \in G\) must be admitted together,
\begin{equation}
z_i = z_{i'}, \qquad \forall i, i' \in G.
\label{eq:squiro_gang_admission}
\end{equation}

\begin{figure}
\centering
\includegraphics[width=0.5\textwidth]{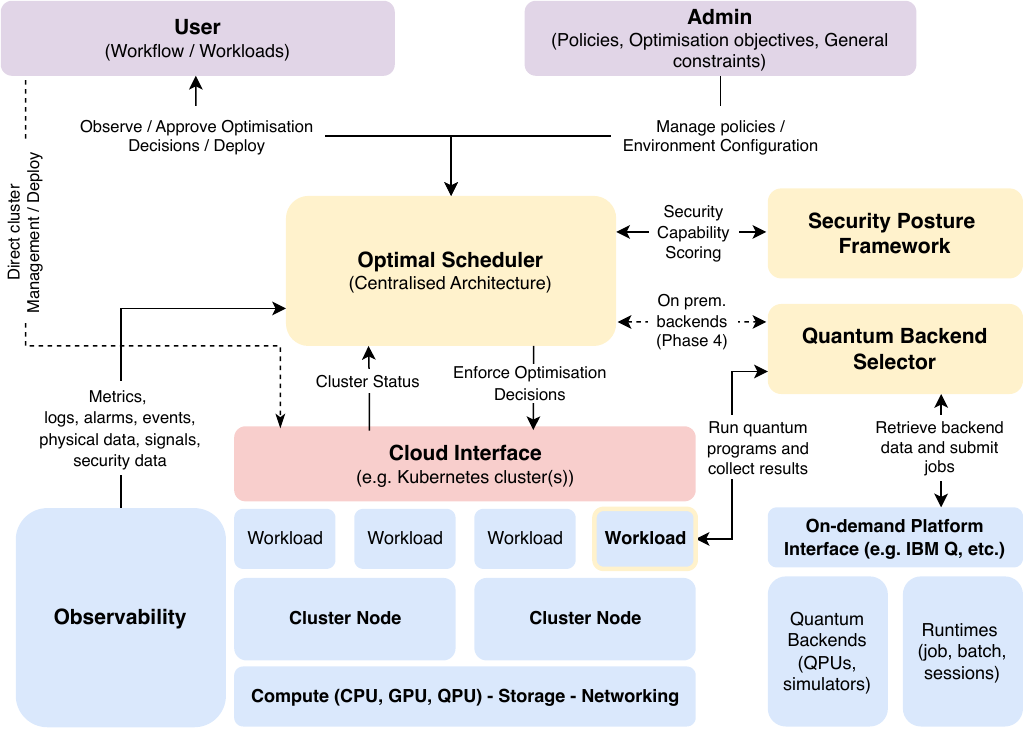}
\caption{SQUIRO centralised optimisation architecture.
Kubernetes supplies workload and cluster state; external modules supply
quantum backend metadata and security evidence; SQUIRO normalises inputs
into the USM and returns placement, validation, or
backend-selection decisions.}
\label{fig:squiro_architecture}
\end{figure}

Let \(\kappa_j\) be the monetary cost rate of node \(j\), \(p_j\) its
active power draw, and \(T\) the accounting interval:
\begin{align}
C(y) &= \sum_{j \in R} \kappa_j y_j, \label{eq:squiro_cost} \\
E(y) &= T \sum_{j \in R} p_j y_j. \label{eq:squiro_energy}
\end{align}
The policy-governed scalar objective is
\begin{equation}
\min \;\alpha C(y) + \beta E(y) + \lambda R(x) + \delta M(x)
- \gamma A(z),
\label{eq:squiro_scalar_objective}
\end{equation}
where \(A(z) = \sum_i \pi_i z_i\) rewards admission by workload priority
\(\pi_i\), \(M(x)\) captures locality, disruption, or fragmentation
penalties (not yet implemented, left for future formalisation), and \(R(x)\) is the residual risk term defined in
Section~\ref{subsec:squiro_risk_framework}.
The current CP-SAT prototype uses a uniform admission reward, with \(\pi_i\) held constant across workloads; per workload priority weighting is the general formulation above.
The weights
\(\alpha, \beta, \lambda, \delta, \gamma\) are policy parameters; they
are not fixed in the formulation but chosen to reflect the operator's
deployment objective. Representative profiles include:
\begin{itemize}
\item \emph{Performance:} maximise admission by setting \(\alpha = \beta
= \lambda = 0\) and keeping \(\gamma\) dominant;
\item \emph{Green:} minimise active-node energy by making \(\beta\)
dominant;
\item \emph{Regulated:} enforce security thresholds via the hard mask
\(a^s_{ij}\) (described in detail in Section~\ref{subsec:squiro_risk_framework}), then minimise
residual risk among feasible placements by setting \(\lambda > 0\).
\end{itemize}
Setting $\gamma$ so that $\gamma\pi_i$ exceeds the maximum attainable cost and energy saving from rejecting any single workload ensures that consolidation is never achieved through arbitrary rejection.

\subsubsection*{Architecture and Control Boundary}
\label{subsec:squiro_architecture}

\begin{figure}
\centering
\includegraphics[width=\columnwidth]{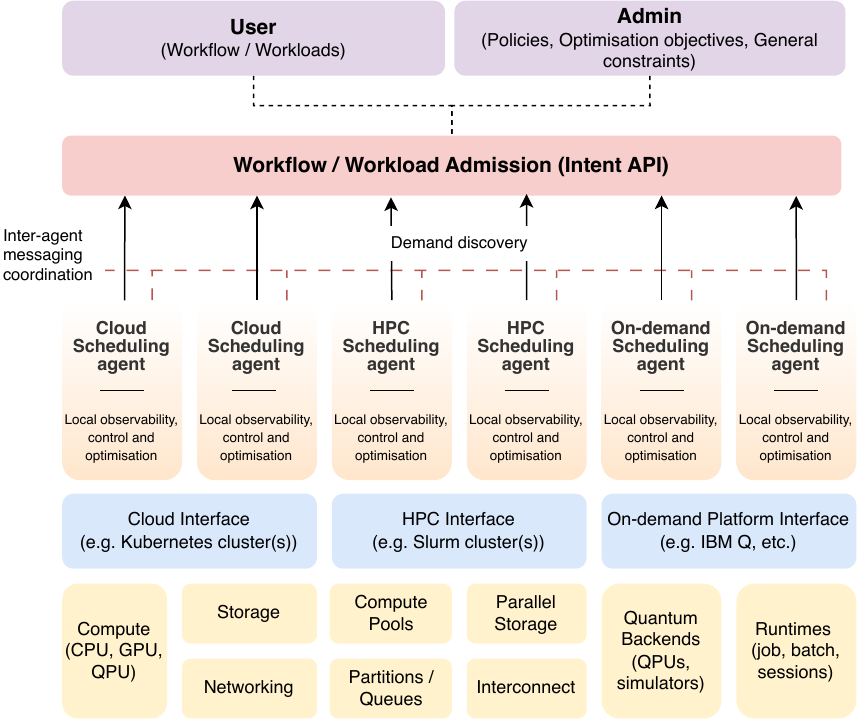}
\caption{Proposed SQUIRO decentralised architecture variant (roadmap design, not yet evaluated). Multiple scheduling agents
each observe a resource partition; the same USM is adapted and instantiated per agent, with
shared policy parameters and lightweight coordination. This also covers the proposed extension to HPC clusters.
}
\label{fig:decentralised_architecture}
\end{figure}

A single Kubernetes administrative domain with homogeneous state
visibility calls for a centralised control architecture, as in Figure~\ref{fig:squiro_architecture}: one logical optimisation plane
observes Kubernetes state and external quantum-provider telemetry,
constructs \(\mathcal{I}_t\), evaluates feasibility and objectives, and
returns a placement or recommendation. In more complex scenarios, such as multi-cluster deployments, a decentralised variant may be appropriate, as shown in
Figure~\ref{fig:decentralised_architecture}.

The decision workflow, shared by both integration modes, is organised in
a fixed sequence. Kubernetes and external adapters first expose a
consistent snapshot: pending workload descriptors, node capacity and
labels, cost and power signals, quantum backend telemetry, and security
evidence. The normalisation layer compiles these inputs into
\(\mathcal{I}_t\), after which feasibility modules compute two
compatibility masks within the same instance: \(a^k_{ij}\) for
Kubernetes node constraints and \(a^s_{ij}\) for security posture.
Quantum backend eligibility (\(a^q_{qb}\)) is instead computed on
demand, only when a workload requests it directly through the client
library (Section~\ref{subsec:squiro_quantum_interface}), rather than
precomputed into \(\mathcal{I}_t\); the two paths converge once
Phase~4 integrates on-premises QPU state directly into the scheduling
instance.
The constraint-programming solver or backend selector then
produces either a placement plan or a ranked backend recommendation.
Finally, the actuation layer translates the result into the mode
permitted by \(\mathcal{A}\), ranging from an offline analytical report
through an advisory recommendation to a direct pod binding. The control
boundary is intentionally narrow: actuation involving stateful or
controller-managed workloads is restricted before solver output reaches
the Kubernetes API.

Two architectural extensions integrate with this pipeline alongside the
core CP-SAT solver. The \emph{security posture framework}
(Section~\ref{subsec:squiro_risk_framework}) processes attestation
evidence and node metadata to produce the security score \(S_j\) and
hard feasibility mask \(a^s_{ij}\); this runs before the solver and
determines which node-workload pairs are admissible. The \emph{quantum
backend selector} (Section~\ref{subsec:squiro_quantum_interface})
processes backend telemetry and circuit-layer workload requirements to
produce the backend eligibility mask \(a^q_{qb}\) and ranked score
determining which backends are eligible and in what
order they should be preferred.
Both modules operate on the same descriptor inputs. The security posture framework feeds its outputs directly into the scheduling instance \(\mathcal{I}_t\); the quantum backend selector runs as a separate callable service and joins \(\mathcal{I}_t\) at Phase~4, as described above. Neither extension alters the abstract model or solver structure.

\subsubsection*{Solver Strategy}
\label{subsec:squiro_solver_strategies}

We use Google OR-Tools CP-SAT \cite{googleCpSat} because the induced problem
and the centralised architecture have a Boolean structure that fits it directly.
CP-SAT is a constraint-programming satisfiability solver:
it handles Boolean placement, activation, and admission variables; 
integer capacity constraints; and compatibility masks through a combination of 
propagation and tree search, with configurable time budgets and parallel workers. 
Where an interpretable linear relaxation is preferred, an ILP formulation over 
the same variable set provides a drop-in alternative. A greedy filter-and-score 
baseline that processes one workload at a time serves as the comparison point in 
the evaluation, making the cost of losing global visibility explicit.
Rolling-horizon partitioning is the proposed mitigation for dynamic arrivals beyond the solver's time budget; topology-dominant
subproblems are a candidate for redirection to min-cost flow; and selected binary subproblems are a candidate for
expression as QUBO instances for quantum annealing \cite{chiavassa2022vnfqa}.
The solver choice follows from the induced problem structure and
architecture, not the reverse.

\emph{Note on the residual risk objective.} The term \(\lambda R(x)\)
is architecturally defined and enters the formulation above, but is not
yet active in the current CP-SAT prototype: the solver enforces the hard
security mask \(a^s_{ij}\) and optimises cost, energy, and admission.
Activating \(\lambda R(x)\) requires integrating the security fit scores
(Section~\ref{subsec:squiro_risk_framework}) into the integer objective,
which is the immediate next implementation step.
The experiment E1 (Section~\ref{sec:evaluation}) 
previews this activation in a separate CP-SAT formulation, validating that the resulting 
risk gradient is informative.

\emph{Remark.} SQUIRO's current implementation targets infrastructure-level
allocation through two separate paths: the CP-SAT solver decides which node a
workload runs on and the global cost-energy-risk trade-off, while the
callable backend selector separately ranks which quantum backend a workload
targets. It does not perform
circuit cutting, compiler-level qubit mapping, or device-level shot
scheduling as these operations produce descriptors and constraints that
SQUIRO consumes, not replaces.
The general model retains workflow edges,
start times, and edge realisation for future workflow-aware extensions;
the current restriction to \(x_k = (z, \mu)\) reflects the Phase~1--2
scope.

\subsection{Quantum Backend Selection}
\label{subsec:squiro_quantum_interface}

\subsubsection*{Overview and motivation}

Selecting a quantum backend is not a static hardware choice. A backend
with the lowest median two-qubit error rate may be the wrong selection for
a given circuit at a given moment: if its queue delay is two hours, a
closed-loop variational algorithm requiring sub-minute latency cannot use
it; if the circuit's estimated execution time exceeds the backend's
coherence time \(T_2\), gates accumulate decoherence errors that aggregate
error metrics do not capture; if the circuit requires all-to-all
connectivity and the backend is a heavy-hex lattice, transpilation overhead
silently degrades effective fidelity. The appropriate backend depends
jointly on the circuit's structure, the workload's execution mode, and the
backend's current operational state.

SQUIRO addresses this through a two-stage selector. The first stage
applies hard feasibility filters: a backend that cannot satisfy the
workload's qubit count, required capabilities, colocation class, or
calibration freshness threshold for its execution mode is excluded
unconditionally.
These are mandatory policy or capability requirements, not preferences: a
backend that fails them is excluded regardless of its
aggregate quality score.
 The second stage ranks the remaining candidates
by a composite score that balances hardware quality against execution
latency, queue pressure, compilation compatibility, and capability fit.
The key quality term is circuit-specific: rather than treating gate error
as a property of the backend in isolation, SQUIRO computes a coherence
quality factor that penalises backends whose decoherence timescale is
shorter than the circuit's estimated execution time, capturing a failure
mode invisible to error-rate ranking alone.

In the abstract model, quantum tasks are workload vertices in \(G_W\)
and quantum backends, simulators, or future integrated QPUs are resource
vertices in \(G_R\), inheriting the same feasibility and objective
structure as classical node placement.
The backend assignment variable
\(u_{qb}\) is designed to be jointly optimised with classical placement
under a unified scheduling instance, a target realised at Phase~4 once
on-premises QPU state becomes locally schedulable. In the current
phase, since no standard Kubernetes primitive exists for
quantum-classical workloads, SQUIRO exposes the backend selector as a
callable service through a provider-agnostic library that
abstracts multiple backends behind a common interface, accepting
circuit-layer information (qubit demand, gate depth, execution mode,
error budgets) as input and returning a ranked list of feasible backends.

\subsubsection*{Descriptors and Hard Feasibility}

Five descriptor types govern backend selection, spanning workload
requirements, hybrid execution coupling, backend hardware, calibration
state, and colocation constraints. Circuit-level parameters such as
qubit demand, gate depth, and execution mode can be provided directly in
the workload descriptor or derived from a resource estimation step using
a provider or compiler analysis step, where available, enabling automatic population of
the feasibility mask without manual specification.

\begin{itemize}
\item \textit{Quantum workload.} Specifies execution mode (\textsc{batch}, \textsc{closed-loop},
\textsc{near-time}, \textsc{real-time}), backend family, qubit demand,
circuit depth and gate counts, and required capabilities; these populate the backend
feasibility mask \(a^q_{qb}\) and the circuit-dependent quality penalties.

\item \textit{Hybrid execution demand.} Couples the quantum task to its classical context via latency
budget, data locality requirement, estimated runtime per shot, and classical preprocessing
and postprocessing demands, governing queue tolerance and locality policy.

\item \textit{Quantum backend.} Defines the resource vertex in \(G_R\) and capability set \(K_b(t)\),
capturing backend family, qubit count, connectivity class (e.g.\ \textsc{heavy-hex}, \textsc{all-to-all}),
native gate set, and provider metadata.

\item \textit{Backend state.} Holds time-dependent calibration data, such as median gate and readout errors, \(T_1\) and \(T_2\)
relaxation times, calibration age, queue state, and estimated queue delay, forming the time-varying component of \(S_t\)
that drives quality, latency, and availability terms in \(\mathcal{O}\).

\item \textit{Colocation class hierarchy.} One of \textsc{remote}, \textsc{network-near}, \textsc{node-near},
\textsc{controller-tight}; determines execution-mode compatibility, latency penalty,
and the QPU integration path across roadmap phases.
\end{itemize}

Implementing the first stage described above, the backend feasibility
mask \(a^q_{qb} \in \{0,1\}\) encodes non-negotiable constraints: a
backend fails the hard filter when its family or capability set does not
match the workload requirement, when it has insufficient physical qubits,
when its colocation class does not satisfy the execution mode's
requirement, or when its calibration data is too stale for the
requested mode.
Calibration age is assessed against execution-mode-specific
thresholds: real-time execution demands the most recent calibration
data, while batch workloads tolerate older snapshots, reflecting that
stale calibration data makes the scheduler's fidelity estimate less
reliable, not that the hardware's actual fidelity degrades. These thresholds are policy-configurable.

The colocation class encodes the physical proximity and integration
depth between the classical scheduling layer and the quantum execution
unit, forming an ordered hierarchy:
\textsc{remote}\,$\prec$\,\textsc{network-near}\,$\prec$\,%
\textsc{node-near}\,$\prec$\,\textsc{controller-tight}. This hierarchy
reflects the SQUIRO integration roadmap (Figure \ref{fig:roadmap})
and is designed to be backward-compatible: a scheduler operating at an
earlier phase assigns \textsc{remote} or \textsc{network-near} backends
and remains valid as integration depth increases. Current on-demand
cloud backends carry \textsc{remote}; in SQUIRO's mapping, HPC-attached QPUs via Slurm
plugins are \textsc{network-near} or \textsc{node-near}; future
node-level quantum capabilities exposed through device plugins or OS
abstractions
correspond to \textsc{controller-tight}. Execution mode requirements
constrain the minimum acceptable class: real-time workloads require
\textsc{controller-tight}, near-time workloads \textsc{network-near} or
tighter, and batch and closed-loop workloads tolerate \textsc{remote}, with closed-loop's tighter timing requirement enforced instead through calibration freshness (Section~\ref{sec:evaluation}).
The
hierarchy acts as both a capability descriptor and a future-integration
contract without altering the abstract model.

\subsubsection*{Backend Scoring}

Let \(Q \subseteq W\) be the quantum tasks and \(B \subseteq R\) the
candidate backends. The variable \(u_{qb}\) indicates assignment of
quantum task \(q\) to backend \(b\):
\begin{align}
u_{qb} &\leq a^q_{qb}, && \forall q \in Q,\; \forall b \in B, \\
\sum_{b \in B} u_{qb} &= z_q, && \forall q \in Q.
\end{align}

Among feasible backends, the composite score is computed per
workload-backend pair, not per backend in isolation: the circuit depth
and execution mode of the specific quantum task determine the quality
penalty, latency sensitivity, and capability fit, making the ranking an
intrinsically workload-specific operation. The score is:
\begin{equation}
\sigma_{qb} = \bigl(w_Q\, Q_{qb} - w_L\, L_{qb} - w_V\, V_{qb}
- w_C\, C_{qb} + w_F\, F_{qb}\bigr) \cdot \rho_q,
\label{eq:squiro_backend_score}
\end{equation}
where \(\rho_q\) is a workload-priority multiplier and the weights
\(w_Q, w_L, w_V, w_C, w_F \geq 0\) are policy-configurable. Each term
reflects a distinct scheduling concern: hardware quality \(Q_{qb}\),
execution latency penalty \(L_{qb}\), queue pressure penalty \(V_{qb}\),
compilation compatibility penalty \(C_{qb}\), and capability fit bonus
\(F_{qb}\). Positive terms contribute to the score; penalty terms reduce
it. The weighting allows operators to reflect deployment priorities: a
research workload tolerant of queue delay benefits from a
quality-dominant profile; a latency-critical iterative algorithm
requires a penalty-dominant profile that demotes saturated or distant
backends.

The hardware quality score \(Q_{qb}\) decomposes into circuit-aware
terms:
\begin{equation}
Q_{qb} =  \max\!\left(0,\; c_1(1 - e_{2q}) + c_2(1 - e_{ro})
+ c_3\, q_\mathrm{coh} - c_4\, p_\mathrm{noise}\right)
\label{eq:quality_decomposition}
\end{equation}
where \(e_{2q}\) and \(e_{ro}\) are median two-qubit and readout error
rates, \(c_1,\ldots,c_4 \geq 0\) are coefficients summing to one, and
\(Q_{qb} \in [0,1]\) by construction.
The first two terms reward low
gate and readout error in the conventional sense. The third term is the
coherence quality factor:
\begin{equation}
q_\mathrm{coh} = \max\!\left(0,\; 1 - \frac{d_{1q} t_{1q} + d_{2q} t_{2q}}
{\min(T_1, T_2)}\right),
\label{eq:coherence_quality}
\end{equation}
where \(d_{1q}, d_{2q}\) are the
circuit depths considering sequential 1- and 2-qubit gates,
\(t_{1q}, t_{2q}\) the corresponding median gate times, and \(T_1,
T_2\) the hardware relaxation and dephasing times. The ratio
\((d_{1q}t_{1q} + d_{2q}t_{2q})/\min(T_1,T_2)\) is the fraction of
the coherence budget consumed by the circuit; \(q_\mathrm{coh}\)
evaluates to zero when the circuit time exceeds \(\min(T_1,T_2)\).
This coherence-margin score is a heuristic ranking proxy, not a validity test: a circuit whose execution time exceeds \(\min(T_1,T_2)\) is unlikely to complete coherently on that backend, independent of its gate error metrics, though the depth-time approximation does not model transpilation, parallel gate execution, dynamical decoupling, or error mitigation.
The term still makes the selection sensitive to a
failure mode that aggregate error statistics cannot detect: a backend
with excellent median gate error but short coherence time is correctly
penalised for deep circuits.

The depth-weighted noise accumulation
\(p_\mathrm{noise} = \min\!\bigl(1,\; n_{2q} e_{2q} + n_m e_{ro}\bigr)\)
accounts for gate and readout error accumulation across the full
circuit, where \(n_{2q}\) is the two-qubit gate count and \(n_m\) the
measurement count.

The remaining terms capture complementary scheduling concerns.
The latency penalty \(L_{qb}\) grows with queue delay and colocation
distance, scaled by an execution-mode factor that is highest for real-time
and lowest for batch, reflecting that iterative quantum-classical algorithms are latency-critical in a way that offline batch jobs are not.
The queue pressure penalty \(V_{qb}\) penalises saturated or heavily
loaded backends to avoid concentrating workload on busy providers. The
compilation penalty \(C_{qb}\) penalises topology mismatch, missing
native gate support, and connectivity conflicts that would increase
transpilation overhead and reduce effective fidelity. The fit bonus
\(F_{qb}\) rewards backends that exceed the feasibility threshold in
positive ways, such as error-mitigation support, preferred colocation
class, or error-correction capability.

SQUIRO returns backend recommendations as proposals that can be inspected
before submission: provider-side execution is governed by the provider
interface, and backend selection does not imply submission authority.

\subsection{Security Posture Framework}
\label{subsec:squiro_risk_framework}

\subsubsection*{Overview and motivation}

The quantum-classical threat model extends classical security concerns with new attack surfaces. The model separates binary requirements from gradational preferences.
Some are absolute: a workload requiring post-quantum key encapsulation
cannot run on a node without it, regardless of cost. This is a
correctness requirement, not merely a suboptimal placement. Others
admit levels: among nodes satisfying all hard requirements, one may
have fresher attestation evidence, stronger cryptographic agility, or
lower side-channel exposure than another, and the more secure host is
preferable though not required.

Standard cloud schedulers conflate these two categories: security checks
are either absent or treated as soft scoring terms that can be traded
against cost.
SQUIRO enforces the distinction explicitly, using the same
two-stage pattern as the backend selector. Hard feasibility first: a
security mask \(a^s_{ij}\) is set to zero if any mandatory requirement of
workload \(i\) is unmet on node \(j\); the solver never considers an
infeasible pair regardless of its objective value. Scored optimisation
second: among feasible placements, quantitative residual risk enters the
objective as a soft term \(\lambda R(x)\), driving placement towards the
most secure adequate host without overriding feasibility.

The framework covers three groups of security dimensions, motivated by a threat model specific to quantum-classical infrastructure. Classical threats persist unchanged. The quantum computing context adds new dimensions.
Harvest-now/decrypt-later attacks make PQC posture
a concern at scheduling time, not only at key-management time.
Multi-tenant QPU access introduces side-channel risks without classical analogues: co-scheduled workloads may infer circuit structure or execution timing from
hardware-level observables.
AI and ML components deployed in
quantum-classical pipelines, for error mitigation, circuit compilation,
or scheduling, carry adversarial risks that become infrastructure
concerns when those models run on scheduled nodes.

\subsubsection*{Dimension Model}
A security \emph{dimension} is an independently-scored security attribute
category (e.g.\ transport encryption, PQC readiness, trusted
computing): each node \(j\) receives a score \(S_{jh}\in[0,100]\) per
dimension \(h\), and dimension scores combine into an overall posture
score via the workload-configurable weighting of
Eq.~\eqref{eq:squiro_posture_score}.
The framework organises multiple dimensions across three groups: 1) \textit{Infrastructure and Compliance}, 2) \textit{Quantum-aware controls}, 3) \textit{AI-driven risk}. Trusted computing
and PQC are the most directly enforced dimensions
in the current implementation, addressing the highest-impact risks for
regulated quantum-classical workloads: hardware-attested platform
integrity and algorithmic resistance to quantum adversaries,
respectively.

\paragraph{Group 1: Infrastructure and compliance.}
Dimensions belonging to this group score classical security capabilities
and general system hardening. Examples are:
\begin{itemize}
    \item \emph{Transport} evaluates the encryption posture of the node's
Kubernetes communication channels, grounded in NIST CSF 2.0 and
SP~800-53 \cite{nistCsf2,nistSp80053}.
The score covers control-plane, pod-network, and management channels.
\emph{Control-plane} covers the minimum TLS version
enforced on the API server and kubelet endpoints and whether mutual TLS
is required for control-plane authentication. The \emph{pod network
layer} assesses whether pod-to-pod traffic is encrypted at the
infrastructure level via the CNI plugin (e.g.\ WireGuard or
IPsec-based overlays). \emph{Management access} channels, SSH and
equivalent remote administration, are scored by authentication mode
and protocol version. These are node-level configuration facts
extractable from attestation evidence or Kubernetes API state; the
dimension assesses what the node's infrastructure enforces, not what
individual services elect to use.
    \item \emph{Hardening} assesses how well the node resists known
attack paths: it considers patch cadence and vulnerability exposure, the
depth of OS and network segmentation, and whether runtime controls such as
endpoint detection and privileged-access authentication are in place.
    \item \emph{Trusted computing} evaluates the integrity evidence a
node can provide. The assurance level depends on the TEE technology
present, the TPM version, whether boot
measurements are secured and reported, and how recently attestation
evidence was renewed. Attestation freshness is a
hard constraint because current integrity cannot be established within the required freshness bound once evidence goes stale: a node that has not renewed its attestation within
the workload's bound is treated as infeasible.
\end{itemize}

\paragraph{Group 2: Quantum-aware controls.}
This group focuses on quantum emerging threats and security controls:
\begin{itemize}
    \item \emph{Cryptography and PQC} \cite{nistPqcStandards,nistPqcMigration}
assesses both classical cryptographic strength and the node's
readiness for the post-quantum transition: whether it supports
NIST-standardised post-quantum key encapsulation mechanisms (KEMs) and
signature schemes, what NIST PQC security level it achieves, and whether
it can operate in hybrid mode and migrate between algorithms as standards
evolve. A node that cannot satisfy a workload's PQC
requirement is infeasible regardless of its classical posture: the
inability to use a post-quantum algorithm during infrastructure management communication cannot be compensated by good patch cadence.
\item \emph{Distributed quantum} covers the QPU control plane and
runtime observability, including control-plane isolation and access
governance, and continuous side-channel monitoring for shared QPU
access~\cite{erata2024quantum}, independent of the cryptographic and
trusted-execution dimensions covered before.
\end{itemize}

\paragraph{Group 3: AI-driven risk.}

\emph{AI security} \cite{nistAiRmf,owaspLlmTop10} is included because ML
components are now embedded throughout classical and quantum-classical
scheduling pipelines alike, circuit compilation and error mitigation on
the quantum side, autoscaling and placement heuristics on the classical
side, and their adversarial risks (data poisoning, supply-chain
compromise, prompt injection in orchestration layers) become
infrastructure concerns once those models run on scheduled nodes. In the
current implementation this group of dimensions is a declaratory compliance check: a node is AI-security-capable if the operator declares documented alignment with an established AI risk framework (ISO/IEC~42001, NIST AI
RMF~\cite{nistAiRmf}, OWASP LLM Top~10~\cite{owaspLlmTop10}), which are voluntary frameworks that do not themselves certify a node.
This is binary: the declared compliance either exists and is current, or it
does not, and a workload that mandates AI governance coverage treats
undocumented nodes as infeasible.
Future work extends this to
\emph{confidential AI execution}: running ML inference and training
inside TEE-protected environments (Intel TDX, AMD SEV-SNP, ARM~CCA) with
attestation of model identity alongside platform integrity, connecting
directly to the trusted computing dimension.

To clarify how constraints operate within a dimension, consider
PQC, which exhibits both types. Whether the node
supports any NIST-standardised PQC KEM at all is
a binary check: absence makes the pair \((i,j)\) infeasible regardless
of all other scores. The NIST PQC security level achieved
(Level~1, 3, or~5) is gradable: a workload requiring Level~3 can be
placed on a Level~5 node, which scores higher, but not on a Level~1
node, which fails the floor check.

\subsubsection*{Hard Feasibility}

The security feasibility mask \(a^s_{ij} \in \{0,1\}\) is zero if any
mandatory requirement of workload \(i\) is unmet on node \(j\):
\begin{equation}
x_{ij} \leq a^s_{ij}, \qquad \forall i \in W,\; \forall j \in R.
\label{eq:squiro_security_compatibility}
\end{equation}
Hard constraints span all three groups. At the network and
cryptographic layer, a node fails when its TLS version falls below the
required minimum, when a required PQC KEM or signature algorithm is
absent, or when the quantum-classical link encryption mode falls short of
the required class. At the trusted computing and quantum control layer, a
missing or insufficiently assured TEE, attestation evidence older than the
workload's bound, absent QPU command signing, or a compliance level below
the required floor each render the node infeasible, as does AI governance
coverage below the workload's threshold.
Altogether the framework enforces
over 50 individual constraint checks across all dimensions (Section~\ref{sec:evaluation}'s E1 evaluates a curated subset of ten), and a
single failed check makes the pair \((i,j)\) infeasible regardless of
objective value.

An additional hard failure is recorded if the node's overall security
score falls below the workload's required floor:
\begin{equation}
S_j < \mathrm{min\_score}_i \;\Rightarrow\; a^s_{ij} = 0.
\label{eq:score_floor_fail}
\end{equation}

\subsubsection*{Posture Scoring}

For each node, the framework computes a posture score
\begin{equation}
S_j = \sum_{h \in H} \omega_h\, S_{jh},
\label{eq:squiro_posture_score}
\end{equation}
where \(H\) is the set of security dimensions, \(S_{jh} \in [0,100]\) is
the dimension score, and \(\omega_h\) is the workload-configurable
inter-dimension weight (\(\sum_h \omega_h = 1\)). Each
\(S_{jh}\) aggregates intra-dimension sub-factors (for example, tunnel
strength and TLS version for transport; TEE assurance level and
attestation freshness for trusted computing; command signing and link
encryption mode for distributed quantum), normalised to \([0,1]\) and
scaled to \([0,100]\).
The overall score maps to six operator-facing heuristic labels, not NIST assurance levels: Inadequate \([0,20)\), Basic \([20,40)\), Moderate \([40,60)\),
Strong \([60,75)\), High-assurance \([75,90)\), and Maximum-assurance
\([90,100]\).

\subsubsection*{Workload-Aware Fit Score and Residual Risk}
Once feasibility is established, the framework distinguishes between a
node's absolute security posture and its specific fit for the workload's
requirements. A node with an overall Maximum-assurance posture may still be a
poor fit for a workload that requires a capability the node lacks or
does not meet at the required level. The fit score captures this
workload-specific dimension.

For each feasible pair \((i,j)\), a capability match score \(M_{ij}\)
is computed by applying requirement-coverage functions against the
workload's specific requirements and combining them with the workload-configured
weights \(\omega_h\). 
For each dimension $h \in H$, a per-dimension requirement-coverage
score $m_{ijh} \in [0, 100]$ is computed by starting from $100$ and
subtracting penalties for each unmet workload requirement in that
dimension. Each $m_{ijh}$ is clamped to
$[0,100]$. The capability match score aggregates these
per-dimension scores using the same workload-configured weights
$\omega_h$ as the posture score:
\begin{equation}
M_{ij} = \sum_{h \in H} \omega_h\, m_{ijh}.
\label{eq:capability_match}
\end{equation}
$M_{ij}$ differs from $S_j$ in what is being weighted, not in whether the weights vary by workload: $S_j$ aggregates the node's raw per-dimension scores $S_{jh}$, while $M_{ij}$ aggregates workload $i$'s per-dimension requirement-coverage scores $m_{ijh}$, both under the same workload-configured weights $\omega_h$.

The workload-aware fit score is
\begin{equation}
\Phi_{ij} = \max\!\bigl(0,\;\min\!\bigl(100,\;\nu_M M_{ij} + \nu_S S_j - P_{ij}\bigr)\bigr),
\label{eq:squiro_security_fit}
\end{equation}
where \(\nu_M\) and \(\nu_S\) are policy-configurable weights
(\(\nu_M + \nu_S = 1\)) balancing workload-specific capability match
against absolute node posture. The demand-pressure penalty
\begin{equation}
P_{ij} = \theta \cdot \max\bigl(0,\; D_i - S_j\bigr)
\label{eq:demand_pressure_penalty}
\end{equation}
applies when the node's overall posture \(S_j\) falls below the
workload's estimated demand score \(D_i\), with \(\theta
\geq 0\) a configurable penalty rate. The demand score is
\begin{equation}
D_i = \sum_{h \in H} \omega_h\, \delta_{ih},
\label{eq:squiro_demand_score}
\end{equation}
where \(\delta_{ih} \in [0,100]\) is workload \(i\)'s required
stringency level in dimension \(h\) (zero where the workload has no
specific requirement in that dimension), normalised on the same scale
as the node score \(S_{jh}\); \(D_i\) therefore lies in \([0,100]\) by
construction, using the same weights \(\omega_h\) as the posture
score. For a typical regulated workload, PQC, attestation, compliance,
and AI-security requirements dominate the sum.

\emph{Note on the demand score.} Eq.~\eqref{eq:squiro_demand_score} states the general weighted-sum formulation. The current prototype instead computes \(D_i\) from an additive, per-requirement scoring pass with a compression factor applied to the portion of the score above a baseline, then clamps the result to \([0,100]\); this keeps the demand scale informative across diverse workload classes without saturating at the maximum too easily. Aligning the implementation with the closed-form weighted sum is future work.

To illustrate: a node achieving $S_j{=}84$ (High-assurance) based on
strong PQC, trusted computing, and transport scores may still be a poor
fit for a regulated workload mandating FIPS~140-3 Level-3 validation if
the node's cryptographic module holds only FIPS~140-2 Level-2 validation.
The capability
match score $M_{ij}$ for the compliance sub-factor evaluates to zero
under this requirement, and $\Phi_{ij}$ falls well below $S_j$. The
same node that is an excellent fit for a research workload without
certification constraints becomes a poor fit for the regulated placement, a mismatch that the absolute posture score $S_j$ alone cannot
detect.

Residual risk is derived from the fit score:
\begin{equation}
r_{ij} = 1 - \frac{\Phi_{ij}}{100}, \qquad
R(x) = \sum_{i \in W} \sum_{j \in R} r_{ij}\, x_{ij}.
\label{eq:squiro_residual_risk}
\end{equation}
\(R(x)\) ranks feasible placements by their residual security exposure;
it does not relax \(a^s_{ij}\). The hard mask and the soft
risk term are complementary: the mask ensures correctness, the risk term
drives optimisation within the feasible region.

\section{Preliminary Evaluation}
\label{sec:evaluation}

\begin{figure*}
\centering
\begin{subfigure}{0.48\linewidth}
  \centering
  \includegraphics[width=\linewidth]{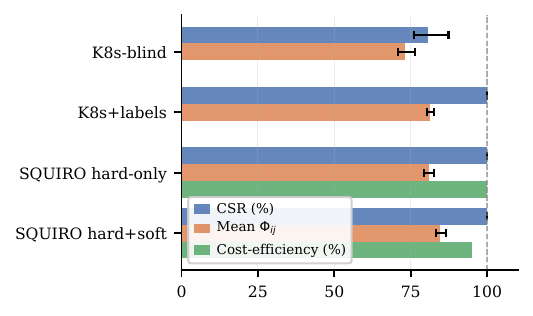}
  \caption{CSR, mean $\Phi_{ij}$, and cost-efficiency across four
scheduling arms (E1, $N{=}30$), normalised to $[0,100]$. The hard
security mask enforces CSR\,=\,1 (\texttt{K8s-blind}: 0.81); admission
is not a differentiator (99--100\% across all arms).
\texttt{K8s+labels} matches \texttt{SQUIRO}'s correctness but not its
efficiency, requiring more than twice as many nodes. Among the \texttt{SQUIRO}
arms, the soft residual-risk term raises mean $\Phi_{ij}$ by 3.7
points at a 3.4\% cost premium.}
\end{subfigure}
\hfill
\begin{subfigure}{0.48\linewidth}
  \centering
  \includegraphics[width=\linewidth]{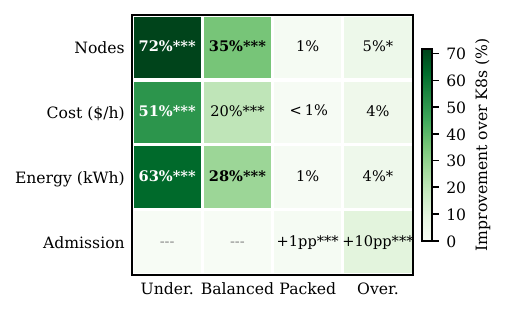}
  \caption{\texttt{CP-SAT-balanced} over \texttt{K8s-greedy} improvement
across four metrics and load regimes (E2, $N{=}30$); green
intensity $\propto$ improvement, admission row shows
\texttt{CP-SAT-performance} improvement ($^{***}p{<}0.001$, $^*p{<}0.05$,
Wilcoxon signed-rank). Global optimisation saves 51\% cost and 63\%
energy in the underloaded regime, 20--28\% in balanced; packed and
overloaded regimes exhaust consolidation headroom, and overloaded
admission improves by 10\,pp.}
\end{subfigure}

\vspace{4pt}

\begin{subfigure}{0.36\linewidth}
  \centering
  \includegraphics[width=\linewidth]{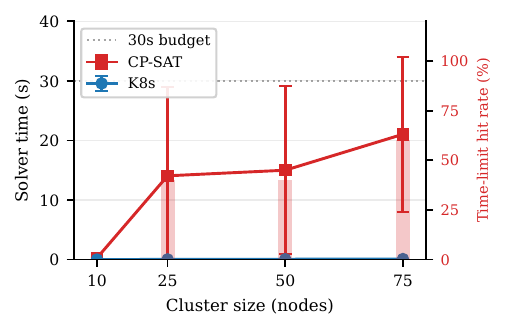}
  \caption{Solver time vs.\ cluster size (E3, $N{=}5$,
mean\,$\pm$1\,std; dotted line\,=\,30\,s budget, red bars\,=\,fraction
of runs exhausting the budget). \texttt{CP-SAT} time grows from 0.27\,s to
21\,s while \texttt{K8s} stays below 100\,ms at all sizes; large
variance at 75 nodes reflects the gap between runs finding optimal
quickly and those hitting the budget.}
\end{subfigure}
\hfill
\begin{subfigure}{0.61\linewidth}
  \centering
  \includegraphics[width=\linewidth]{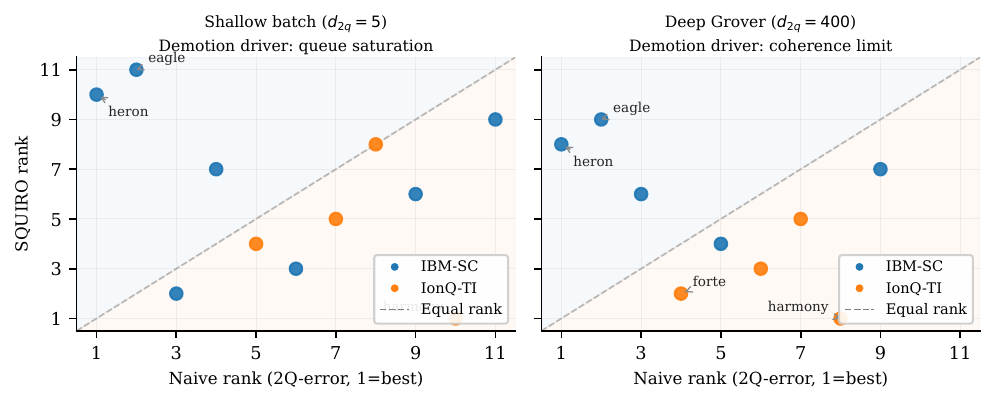}
  \caption{SQUIRO rank vs.\ naive 2Q-error rank for feasible backends
(E4; blue\,=\,IBM superconducting, orange\,=\,IonQ trapped-ion). Left:
shallow batch ($d_{2q}{=}5$), top-naive IBM backends demoted by queue
saturation, \texttt{ionq\_harmony} promoted to rank~1. Right: deep Grover
($d_{2q}{=}400$), \texttt{ibm\_heron} and \texttt{ibm\_eagle} demoted by coherence limit
($q_\mathrm{coh}{=}0$), IonQ backends rise. Points above the diagonal
are demoted, below are promoted; queue saturation and coherence limits
are the two dominant divergence mechanisms between SQUIRO and naive
ranking.}
\end{subfigure}

\caption{Preliminary evaluation results (synthetic clusters).}
\label{fig:eval}
\end{figure*}

The experiments are
conducted on synthetic clusters using simulated hardware parameters
derived from published specifications.
The purpose is to validate the framework structural claims, detailed per experiment below. Performance at scale is
treated as an open challenge rather than a claim; results should be
interpreted as evidence for the framework's structural properties and
do not replace evaluation on a live quantum-classical cluster.

We validate the distinct claims of the SQUIRO framework with four experiments.
\textit{E1} tests whether the hard security mask enforces placement constraints
by construction, whether a native Kubernetes (K8s) encoding of the same mask
matches this guarantee under greedy placement, and whether the
residual-risk term improves placement quality once constraint
satisfaction saturates.
\textit{E2} examines whether global
optimisation produces measurable cost and energy gains over greedy
placement, and how the benefit varies with cluster load.
\textit{E3} identifies the
cluster scale at which CP-SAT solve time becomes the binding operational
constraint.
\textit{E4} tests whether the two-stage backend selector diverges
meaningfully from naive two-qubit-error (2Q-error) ranking and characterises the mechanisms
driving that divergence.

All instances are synthetic, generated from distributions
grounded in published specifications: \texttt{GCP us-central1} on-demand
pricing~\cite{gcpPricing2025}, IBM Quantum calibration
data~\cite{ibmQuantumPlatform}, and IonQ hardware
specifications~\cite{ionqHardwareSpecs}.
All experiments were
executed on a MacBook Pro (Apple M2 Max, 12-core CPU, 32\,GB unified
memory) running \texttt{Python~3.11.4} and \texttt{OR-Tools~9.15} (CP-SAT solver).

\textbf{E1\,--\,Security-aware placement.}
The cluster comprises 50 nodes hosting 150 pods with heterogeneous
security requirements at approximately 55\% mean load. Security
posture uses a correlated compliance model over a curated 10-check
subset of the framework's dimensions, deliberately calibrated so a
genuine, non-degenerate feasible region exists for every workload
profile rather than an all-or-nothing outcome: node pools are sized to
each profile's demand share plus margin for real choice, and research,
enterprise, and regulated pods require 0, 4, and 10 of the 10 checks
respectively.
Significance uses the Wilcoxon signed-rank test (paired,
non-parametric), with effect size reported as rank-biserial $r$.
Four arms are compared: \verb|K8s-blind| (no security model), \verb|K8s+labels|
(the identical hard mask $a^s_{ij}$ encoded via nodeAffinity and
taints, placed greedily), \verb|SQUIRO hard-only| ($\lambda{=}0$), and \verb|SQUIRO hard+soft| ($\lambda{>}0$, previewing the residual-risk term in a
CP-SAT formulation);
the solver runs with a 10\,s time limit and 32 workers.
Fig.~\ref{fig:eval}(a) shows constraint satisfaction rate (CSR), mean fit score $\Phi_{ij}$, and a composite cost-efficiency score: cost and energy are each min-max normalised across the four arms shown (inverted, so lower cost/energy scores higher) and averaged, anchoring the best-observed arm near 100 and the worst near 0.
\verb|K8s-blind| reaches CSR\,${=}\,0.81$ overall against CSR\,${=}\,1.00$ for
all three security-aware arms (Wilcoxon $W{=}0$,
$p{=}1.9{\times}10^{-9}$, $r{=}1.00$); the pooled figure is inflated by
research pods, which carry no requirement, and masks a regulated-only
CSR of just 0.26. Admission is not a differentiator; all arms admit
99--100\% of pods.
\verb|K8s+labels| and \verb|K8s-blind| use an identical node footprint (50 of 50,
same cost and energy, every seed) regardless of whether the hard mask
is enforced: the inefficiency is greedy placement, not security
filtering. SQUIRO needs only 21--23 nodes for the same workload
(Wilcoxon $W{=}0$, $p{=}1.9{\times}10^{-9}$, $r{=}1.00$): native labels
match SQUIRO on correctness but not efficiency.
Activating the soft term raises mean $\Phi_{ij}$ from 81.0 to 84.7
(Wilcoxon $W{=}0$, $p{=}1.9{\times}10^{-9}$, $r{=}1.00$) at a 3.4\%
cost and 4.7\% energy premium: the residual-risk gradient among
feasible placements is informative.

\textbf{E2\,--\,Multi-objective consolidation.}
Clusters contain 20--40 GCP-priced nodes across 4 tiers
(8--64 cores, \$0.27--\$2.14/h) spanning 2 availability zones; workloads
are scaled to four load regimes: underloaded (spare capacity
throughout), balanced (moderate utilisation), packed (near-saturation),
and overloaded (demand exceeds supply).
Three schedulers are compared: \texttt{K8s-greedy}, \texttt{CP-SAT-performance} (maximises admission), and \texttt{CP-SAT-balanced} (joint cost--energy objective); all CP-SAT variants use a 10\,s time limit with 32 workers.
Reported metrics are admission rate, active node count, hourly cost (\$/h),
and energy consumption (kWh).
Fig.~\ref{fig:eval}(b) shows mean hourly cost across load regimes.
In the underloaded regime, \texttt{CP-SAT-balanced} reduces cost by 51\%
(\$9.20 vs.\ \$18.73) and energy by 63\% (5.39\,kWh vs.\ 14.69\,kWh)
by packing all demand onto a fraction of nodes (8.4 vs.\ 29.6 active).
In the balanced regime the savings are 20\% and 28\% respectively.
In packed and overloaded regimes near-total capacity utilisation forces
both schedulers to activate most nodes; consolidation headroom disappears.
The value of global optimisation in the overloaded regime
shifts to admission: \texttt{CP-SAT-performance} admits 9.7~pp more workloads than
\texttt{K8s-greedy} (88.4\% vs.\ 78.7\%).
Wilcoxon tests confirm significance for underloaded and balanced contrasts
($W{=}0$, $p{<}10^{-8}$, $r{=}1.00$, $n{=}30$).

\textbf{E3\,--\,CP-SAT scalability.}
Cluster size ranges from 10 to 75 nodes (up to 400 pods)
across 4 security tiers at a balanced load of 75\%\,$\pm$\,5\%.
\texttt{K8s-greedy} is compared against \texttt{CP-SAT-balanced} under a 30\,s time limit
with 8 workers; reported metrics are solver time and admission rate.
Fig.~\ref{fig:eval}(c) shows solver time as a function of cluster size.
\texttt{CP-SAT} time grows from 0.27\,s ($n{=}10$) to 21\,s ($n{=}75$), with
60\% of 75-node instances exhausting the 30\,s budget; \texttt{K8s} stays below
100\,ms at all sizes.
\texttt{CP-SAT} proves optimality within budget in all \(n{=}10\) runs, but the time-limit hit rate rises to 40\% at \(n{=}25\) and \(n{=}50\), and 60\% at \(n{=}75\). These results motivate evaluation of rolling-horizon partitioning to maintain bounded, predictable solve times beyond the smallest instances, where full optimality cannot always be certified within budget.
This points to
a fundamental operational constraint: a scheduling system must produce
decisions fast enough to match the tempo of cluster dynamics.

\textbf{E4\,--\,Circuit-aware backend selection.}
The backend pool comprises 11 hardware devices: 7 IBM superconducting (\(T_2=45\text{--}150\,\mu\mathrm{s}\)) and 4 IonQ trapped-ion (\(T_2=0.5\text{--}1.2\,\mathrm{s}\)). Ten workload types span three execution modes (batch, closed-loop, and near-time) across three circuit depth classes.
Calibration freshness thresholds follow implementation defaults: 24\,h
for batch, 6\,h for closed-loop, and 2\,h for near-time workloads.
The SQUIRO two-stage selector is compared against naive 2Q-error ranking;
the reported metric is rank position for each backend under each method.
The two-stage selector applies hard feasibility filters (including
mode-specific calibration freshness thresholds) then scores backends
using \(\sigma_{qb}\) (Eq.~\ref{eq:squiro_backend_score}).
Fig.~\ref{fig:eval}(d) plots SQUIRO rank against naive 2Q-error rank
for two contrasting workloads, exposing the two dominant divergence
mechanisms.
For the shallow batch circuit ($d_{2q}{=}5$, left panel):
\texttt{ibm\_heron} and \texttt{ibm\_eagle} carry the best 2Q-error (naive ranks~1--2) but
have saturated queues (7\,200\,s estimated delay); the queue pressure
 penalty \(V_{qb}\) demotes them to SQUIRO ranks~10--11; \texttt{ionq\_harmony}
(naive rank~10) rises to rank~1 because it is idle and its coherence margin 
is scored as adequate for a shallow circuit.
For the deep Grover circuit ($d_{2q}{=}400$, right panel): circuit
execution time ${\approx}160\,\mu$s exceeds $T_2$ of \texttt{ibm\_heron}
($150\,\mu$s) and \texttt{ibm\_eagle} ($140\,\mu$s); \(q_\mathrm{coh}\) reaches
zero and both are demoted to SQUIRO ranks~8--9 by the coherence quality
term; IonQ backends with longer $T_2$ are ranked at the top by the same term.
All near-time workloads find no feasible backend: every provider carries
\textsc{remote} colocation class, which the hard filter excludes for near-time execution by construction; this validates the filter implementation rather than demonstrating the physical necessity of on-premises QPU integration (Phase~4,
Figure \ref{fig:roadmap}).
Closed-loop workloads (\texttt{vqe\_10q}, \texttt{vqe\_deep\_20q}) show the same queue-driven demotion pattern as the shallow-batch case: saturated IBM backends fall from naive ranks 1--2 to SQUIRO ranks 5--7, but fewer than half of the 11 candidate backends remain feasible under the tighter 6\,h calibration-freshness threshold for this mode. This pair is omitted from Fig.~\ref{fig:eval}(d) for space.
Divergence from naive ranking here
demonstrates that the score responds to coherence and queue conditions
as designed; it does not independently validate that the top-ranked
backend yields higher circuit fidelity or shorter time-to-solution in
practice.
Independent outcome validation, for example via noise-model
simulation, is left as future work.

The results expose three operational boundaries. The security mask result (E1) confirms that correctness cannot be
achieved by scoring alone: a scheduler without a security model
satisfies strict security requirements only by chance, dropping to
roughly one in four for regulated workload classes; enforcing the mask
costs nothing in admission or node count when compared fairly against
an equally greedy baseline, so the gap between native labels and
global optimisation is about placement efficiency, not security
overhead. Among equally compliant placements, the soft residual-risk
term still recovers a meaningful quality gain. The
consolidation result (E2) shows that the value of global optimisation
is load-regime-dependent: greatest in underloaded clusters where
consolidation headroom exists, and marginal in packed clusters where it
does not, suggesting that CP-SAT optimisation is most valuable when the
cluster is not saturated, with incremental scheduling as a fallback as
load approaches capacity.
The scalability result (E3) shows the first observed time-budget instability in this configuration near 25 nodes for single-batch CP-SAT, where the time-limit hit rate first becomes non-trivial (40\%); rolling-horizon re-optimisation, replanning the next \(k\) workloads on each event rather than the full pending set, is a candidate mitigation left for future evaluation.

\section{Conclusion and Future Work}
\label{sec:conclusion}

This paper proposed USM and SDM, a platform-independent model and methodology for deriving quantum-classical schedulers that treat security posture and quantum hardware awareness as first-class scheduling concerns. Instantiated over Kubernetes, SQUIRO implements and evaluates
a strict separation between mandatory security requirements and
gradational preferences, a circuit-aware quantum backend selector, and
a colocation class hierarchy spanning today's remote quantum access
and tomorrow's node-level integration.
SQUIRO's portability is intended to support future extensions including HPC/Slurm instantiation applying the
same design methodology to a different deployment target.
Whether this holds at larger scale, across greater backend
heterogeneity, and under stricter security regimes remains open; future development and evaluation will determine whether the proposed model structure is sufficient or whether it requires further refinement.

Results show that treating security as an afterthought is inadequate for hard constraint satisfaction, and that the soft residual-risk term recovers a measurable 3.7-point mean security fit-score gain among otherwise equally compliant placements, at the cost of only a few points of cost-efficiency (E1). The same results also show a considerable advantage for SQUIRO over the Kubernetes baseline, with up to 51\% cost reduction and 63\% energy reduction (E2).

Once the model is extended to treat QPUs as node-level capabilities, via device plugins \cite{kubernetesDevicePlugins} or OS abstractions \cite{ramsauer2025towards}, the on-premises quantum case is expected to exhibit the sharpest novelty of the model: the colocation class will shift from \textsc{remote} to \textsc{controller-tight}, calibration freshness will become locally observable, and queue dynamics will become scheduler-visible.

Future developments will test whether the same USM can represent this shift through new QDMI-compatible adapters
\cite{burgholzer2026qdmi,wille2024qdmi}, QPU capability descriptors,
and shot-allocation capacity models, rather than requiring model revisions.

Finally, the longer-horizon ambition and concrete validations are a live-cluster study and joint node/backend optimisation, en route to the full end-to-end, cross-layer optimisation, as per our roadmap in Figure~\ref{fig:roadmap}.

\section*{Declaration of Generative AI and AI-assisted technologies in the writing process}

During the preparation of this work the authors used Claude (Anthropic),
ChatGPT, and Codex (OpenAI) for editorial assistance. After using these
tools, the authors reviewed and edited the content as needed and take
full responsibility for the content of the published article.

\section*{Declaration of Interest Statement}
The authors are affiliated with Helix 42, which may pursue commercial development of technologies related to the SQUIRO framework described in this work. The authors declare no other competing financial interests or personal relationships that could have appeared to influence the work reported in this paper.

\section*{Funding}
This research did not receive any specific grant from funding agencies in the public, commercial, or not-for-profit sectors.

\section*{Data availability}
The data that support the findings of this study, including the result files underlying every reported figure, are available from the corresponding author upon reasonable request; the experimental tooling itself is not released.

\bibliography{references}

\end{document}